\documentstyle[12pt,aaspp4]{article}
\def\h75{$h^{-1}_{75}$}
\def\deg{\ifmmode^\circ\else$^\circ$\fi }
\def\solar{\ifmmode_{\mathord\odot}\else$_{\mathord\odot}$\fi }
\def\arcs{$''$\ }

\def\app{$\approx$~}

\def\etal{et~al.\ }

\def\simlt{\mathrel{\hbox to 0pt{\lower 3.5pt\hbox{$\mathchar"218$}\hss}
\raise 1.5pt\hbox{$\mathchar"13C$}}}
\def\simgt{\mathrel{\hbox to 0pt{\lower 3.5pt\hbox{$\mathchar"218$}\hss}
\raise 1.5pt\hbox{$\mathchar"13E$}}}

\newcommand{\parcs}{.\hspace{-0.09cm}$''$}
\newcommand{\beq}{\begin{equation}}
\newcommand{\eeq}{\end{equation}}
\slugcomment{Submitted to Astrophysical J. Letters.}

\lefthead{Pinkney, Holtzman, \& Garasi}
\righthead{WFPC2 Observations of the Cooling Flow Elliptical in Abell 1795}

\begin{document}

\title{WFPC2 Observations of the Cooling Flow Elliptical in Abell 1795
\footnote{Based
on observations with the NASA/ESA {\it Hubble Space Telescope}, obtained at
the Space Telescope Science Institute, operated by AURA Inc under contract
to NASA}}

\author{Jason Pinkney\altaffilmark{2}, 
Jon Holtzman,\altaffilmark{2}, 
Christopher Garasi,\altaffilmark{2}, \\
Alan~M.~Watson\altaffilmark{2},
John~S.~Gallagher~III\altaffilmark{4},
Gilda~E.~Ballester\altaffilmark{5},
Christopher~J.~Burrows\altaffilmark{6},
Stefano~Casertano\altaffilmark{6},
John~T.~Clarke\altaffilmark{5},
David~Crisp\altaffilmark{7},
Robin~W.~Evans\altaffilmark{7},
Richard~E.~Griffiths\altaffilmark{8},
J.~Jeff~Hester\altaffilmark{9},
John~G.~Hoessel\altaffilmark{4},
Jeremy~R.~Mould\altaffilmark{3},
Paul~A.~Scowen\altaffilmark{9},
Karl~R.~Stapelfeldt\altaffilmark{7},
John~T.~Trauger\altaffilmark{7},
and
James~A.~Westphal\altaffilmark{10}}
\altaffiltext{2}{Department of Astronomy, New Mexico State University, Dept 4500
Box 30001, Las Cruces, NM 88003}
\altaffiltext{3}{Mount Stromlo and Siding Springs Observatories, Australian National 
University, Private Bag, Weston Creek Post Office, ACT 2611, Australia}
\altaffiltext{4}{Department of Astronomy, University of Wisconsin -- Madison, 475 
N. Charter St., Madison, WI 53706}
\altaffiltext{5}{Department of Atmospheric, Oceanic, and Space Sciences, University 
of Michigan, 2455 Hayward, Ann Arbor, MI 48109}
\altaffiltext{6}{Astrophysics Division, Space Science Department, ESA \& Space 
Telescope Science Institute, 3700 San Martin Drive, Baltimore, MD 21218}
\altaffiltext{7}{Jet Propulsion Laboratory, 4800 Oak Grove Drive, Pasadena, CA 91109}
\altaffiltext{8}{Department of Astronomy, Johns Hopkins University, 3400 N. Charles 
St., Baltimore, MD 21218}
\altaffiltext{9}{Department of Physics and Astronomy, Arizona State University,
Tyler Mall, Tempe, AZ 85287}
\altaffiltext{10}{Division of Geological and Planetary Sciences, California Institute 
of Technology, Pasadena, CA 91125}

\begin{abstract}
We present WFPC2 images of the core of 
the cooling flow cD galaxy in Abell 1795.
An irregular, asymmetric dust lane 
extends 7 \h75 kpc in 
projection to the north-northwest.  The dust shares the morphology
observed in the H$\alpha$ and excess UV emission.
We see both diffuse and knotty blue emission around the dust lane, especially
at the ends.
The dust and emission features lie on the edge of the radio lobes,
suggesting star formation induced by the radio source or the deflection
of the radio jets off of pre-existing dust and gas.
We measure an apparent R$_V$ significantly less than 3.1, implying that the
extinction law is not Galactic in the dust lane, or the presence
of line emission which is proportional to the
extinction.  
The dust mass is at least 2$\times10^{5} h_{75}^{-2}$ M\solar\ and is
more likely to be 6.5$\times10^{5} h_{75}^{-2}$ M\solar.

\end{abstract}

\keywords{galaxies: elliptical and lenticular, cD - galaxies: clusters: general
- dust extinction }

\pagebreak

\section{Introduction}

Abell 1795 
is a richness class 2 cluster and is one 
of the prototypical cooling flow clusters.  Both X-ray spectral 
and surface brightness 
studies suggest an accretion rate of $\sim$ 200 $h^{-2}_{75}$ M\solar/yr 
through a cooling radius of $\sim$ 150 \h75 kpc.  
Like other cooling flow clusters, A1795 contains a cD galaxy at the peak of X-ray 
emission.

Optical observations of cooling-flow dominant galaxies (CFDs)
reveal unusual phenomena possibly related to cooling flows (see reviews
by Baum 1992, and Fabian 1994).
Some contain nebulosity and filaments extending tens of kpc from
the nucleus.
The H$\alpha$+[NII] luminosities for CFDs are much higher than for non-cooling 
flow ellipticals of equivalent radio power and correlate with mass
accretion rate (\cite{HBvBM}).
The emission line ratios suggest that both photo- and shock-ionized
gas may be present.
McNamara \& O'Connell (1992) find anomalous colors in  the inner \app 20 kpc
of CFDs; the bluest colors occur in the strongest cooling flows.
There are also ``blue lobes" of excess UV on top of the widespread blue light 
(McNamara \& O'Connell 1993, hereafter MO93).

There is also evidence for dust in CFDs. 
NGC 1275 and 4696 
have dust patches which are easily resolvable from the ground.
Dust is inferred to be mixed with the emission-line gas
from hydrogen recombination line ratios (Hu 1988), and from the
absence of strong Ca II lines (Donahue \& Voit 1993).
About 50\% of CFDs are detected in far IR, with luminosities suggesting
larger dust masses than those inferred from optical observations (Bregman, McNamara, 
\& O'Connell 1990).  
In Abell 1795, dust has not been clearly revealed
in optical imaging (MO93).
The presence of dust in CFDs is surprising because of the hot 
gas environment which should destroy dust on short timescales.

About 71\% of CFDs contain radio sources which are generally
amorphous or bipolar, and powerful (\cite{Bu90}).  They are confined
to the inner \app 10 kpc of the CFD. This is also the realm of
the emission line nebulosity and blue colors, so it has long
been suggested that the radio source drives these phenomena 
(van Breugal \etal 1984; MO93).
The blue lobes (regions of UV excess) found in A1795, A2597 
and NGC 1275 are observed to lie roughly on the edges of radio lobes
(MO93; \cite{MOS95}, hereafter MOS95).
H$\alpha$ and X-ray emission are also located along
the edges of radio lobes (\cite{CHJY83}, MOS95). 
Photoionization of the H${\alpha}$ nebulae by an AGN is ruled out by 
inspecting the dependence of line ratios on distance from the AGN
(Johnstone \& Fabian 1988).  
Scattered light from the AGN is an unlikely cause of the blue lobes 
because the lobe light is not highly polarized
(McNamara \etal 1996a).
It appears more likely that
hot, young stars produce the blue lobes (MOS95), and perhaps that star 
formation is being induced by the radio
source (De Young 1995).

We present WFPC2 images of the center of the cD in
A1795.
The images clearly reveal a dust lane along with bluish 
filamentary and knotty emission features. 
There is a close correspondence between
the radio source and the dust lane and emission features, and  
we discuss the nature of the radio/dust interaction.
We also estimate the mass
of the dust lane and compare the extinction law with  
the Galactic law.  

\section{The Data}

An integration of 1780 s was obtained in each of the
F555W and F702W filters using the Planetary Camera (PC)
on 24 February 1994. 
We transformed the instrumental magnitudes into Johnson V and R, respectively,
using the synthetic WFPC2 transformation given by Holtzman \etal
(1995).  Galactic reddening is negligible in the direction of A1795.
 
We combined our F702W and F555W images to create a color image (Fig. 1).
We subtracted an elliptical model from each image 
and averaged the results to produce Fig. 2.
A dust lane appears dark in Fig. 2 (labeled {\bf D}) and as a dark,
reddish feature in Fig. 1.
The absorption appears greatest near the nucleus, {\bf N}.  We 
suspect the nucleus is the sharp peak seen at the
same pixel in each filter, although dust may obscure the true nucleus.  
In the southernmost ({\bf SKE}) and northernmost ({\bf NKE})
extent of the dust lane, we see blue diffuse and knotty emission.  
Here, and in general around the periphery of the dust, we see the bluest
knots, unresolvable in ground-based data. 
These may be star clusters, and are discussed in Holtzman et al. (1996).  
Some of these ({\bf T}) are very elongated with the same position angle;
two have unusual trails pointing NNW.
Another notable feature is the southern linear filament ({\bf SLF})
which appears to cross the cD just east of the nucleus. 
A northern filament ({\bf NLF}) appears to be a faint continuation
of the {\bf SLF}. 
The faint H$\alpha$ filaments ({\bf FF}) in the SE corner are the end
of a large filament observed to extend 45\arcs 
south of the nucleus (\cite{CHJY83}).
In general, the image shows numerous clusters and bluish
`stellar debris' (e.g., {\bf SD}).  In a separate {\it Letter}, 
McNamara \etal (1996b) suggest that the {\bf NLF} and {\bf SD}
may be tidal debris from nearby ellipticals {\bf E1} and {\bf E2}. 

The dust morphology is similar to the ``irregular" and ``asymmetric" dust
seen in other HST-observed ellipticals (\cite{vanD95}). 
The dust is co-spatial with an H$\alpha$ feature (\cite{vBr84})
and with the U-band excess (MO93).
The U-band absorption feature in MO93 appears to have a 
different position than the dust in our image.  

The majority of the galaxy, away
from emission and absorption features, has (V-R)$\sim$ 0.5-0.6.  
This is slightly bluer than most dominant ellipticals presented by 
Mackie, Visvanathan \& Carter (1990).
This entire inner region was found to be relatively blue in (U-I)
(MO93).  
The dust lane is red, with an average (V-R)$\sim$ 0.85.
The reddest region has (V-R) = 1.4 and is located within an unusual `cap' 
north of the nucleus ({\bf N} in Fig 2) which
does not show strong dust absorption.
The bluest regions occur in the {\bf SLF} and have V-R $\sim$ 0.0.
Most of the {\bf SLF} has V-R $\sim$ 0.4, but there is a parallel red
region running along the its west side (Fig. 1) -
this filament may contain dust hidden under emission.
When our (V-R) map is smoothed to {FWHM\app 1\parcs0}, it is apparent
that the {\bf SLF} is the bluest region in (V-R).  
This is near the edge of the southern radio lobe and
is coincident with one of the two blue lobes in MO93.
The northern blue lobe in MO93 coincides with the dust
lane and {\bf NKE} (Fig. 2).  It does not stand out in our smoothed
color map because the blue knots are mixed with red dust.

\section{Coincidence with Radio Source}
 
In Fig. 3, we overlay contours of 3.6 cm radio 
emission (Ge \& Owen 1993).
We assume that the radio core aligns with the apparent optical 
nucleus.
We see a high degree of correspondence between the radio
emission and optical features.  
The {\bf SLF} shows its brightest knots where it meets 
the radio lobe ({\bf SKE}). 
There appear to be parallel rows of
emission and star formation in this filament
(some inside of the radio emission), as if
material was plowed by an advancing radio lobe.
In general, the southern lobe appears bounded by the dark dust lane
and the southern filament, the west side of which appears to be
traced by dust (reddish in Fig. 1).
Similarly, the northern radio lobe is nestled in the 
arc-shaped dust lane.  Midway along the lane, the lobe appears 
to overlap the dust, but in general, the radio plasma
appears bounded by the dust lane.
A physical contact is supported by the sharper 
gradients in radio emission on the sides of the radio lobes which
are next to the dust lane or filament.  This is more apparent
on the 6cm map of van Breugal \etal (1984) which
shows more extended emission on the east side of the
northern lobe, and to the south and west of the southern
lobe.
\nopagebreak
\section{The Dust Lane}

The high resolution of these data 
allows us to map the absorption in A1795 with precision.
We modeled the starlight of the underlying elliptical using
ELLIPSE in IRAF\footnote{IRAF is distributed by the
National Optical Astronomy Observatories, which is operated by the
Association of Universities for Research in Astronomy, Inc.
(AURA), under cooperative agreement with the National Science 
Foundation.}.
To compensate for the presence of dust
and excess emission,  we excluded these regions from the fit; in 
F702W, we filled in excluded regions assuming reflection symmetry.
The ellipse
fitting was done by first fitting the outer 45 to 350 pixels 
(2$''$ to 15\parcs9) holding only the center fixed on the apparent 
nucleus, 
and then fitting the inner 45 pixels holding the center, position
angle and ellipticity fixed at values determined from the
innermost annuli of the outer fit.
If we fit for the center with the outer 
isophotes we get a location about 0\parcs3 south
of the central peak, so the dust may be obscuring the nucleus.
The quality of the fits was judged by plotting 3-pixel-wide slices 
along three, apparently dust-free radial paths in the model-divided
image.  The photometric errors are $\sim$ 3\% and the model always matches within 10\%.  

We divided the data by the models to create extinction
maps.  The average extinction in 210 $3\times 3$ pixel regions 
is shown for both filters in Fig. \ref{ext}.  
The largest extinctions are 0.55 and 0.77 mag for A$_{702}$ and 
A$_{555}$, respectively.  The dashed lines outline the region
that the extinctions can occupy for a physically thin sheet of dust obeying
a Galactic extinction law (\cite{CCM89}).  
Our data appear outside of this region, obeying a linear relation for 
$A_{702}<0.3$ given by  
$A_{555}=(1.62\pm0.05)A_{702} +0.039\pm0.006$.  
The observed slope is greater than the maximum
expected for Galactic extinction, 1.323, at the 5.9$\sigma$ level.
The intercept is significantly different from 0.0, and is
inconsistent with any extinction law.

We investigated possible systematic errors in our elliptical models
which could reproduce our observations.
Multiplicative errors in our models can only change the intercept.
Our intercept is consistent
with the F702W model being too small by 2.7\%, or the F555W model
being too large by 3.7\%.  Uniform additive errors can change both the
intercept and the slope.  An additive error of 0.6 DN
in F702W can reproduce our intercept,
but the slope is still $\sim$ 1.6
after this correction.  To bring the slope to the Galactic value
would require implausibly large additive errors.

Many natural phenomena make the apparent slope {\em less} than the 
true slope. 
Goudfrooij et al. (1994b, hereafter GDHN) show that foreground 
starlight, an homogeneous mixture of stars and dust, and scattered 
light all decrease the observed slope compared to
the true extinction law.
In fact, we see the influence of foreground light in Fig. \ref{ext} as a `bending
over' of the points at high extinctions,
and we have attempted to compensate by fitting to low-extinction points only.

We find that both the apparent non-Galactic slope and non-zero intercept 
may be consequences of emission-line gas mixed with the dust.
The H$\alpha$+[NII] map of A1795 (van Breugal \etal 1984) 
shows emission co-spatial with the dust which will decrease
the apparent extinction in F702W and cause a positive intercept.
However, if excess emission exists in the F555W band, it moves the  
intercept in the opposite sense and can cancel the F702W effect.
We summed the equivalent widths of all emission lines
in the ``central nebula" using the data of Anton (1993) and
find a {\em net contribution} of 4.0\% to F702W.
Thus, the excess emission in F702W is enough to 
produce the observed intercept.

The emission lines can also account for our non-Galactic slope
if the emission is proportional to the extinction.
Such a proportionality is observed in NGC 4696 (Sparks \etal 1989, hereafter SMG),
and is expected if the position of dust is correlated with that of
gas.
To produce the observed slope and intercept from an intrinsically
Galactic extinction law, 
a total of 4031 DN in emission is required.
The H$\alpha$+[NII] flux measured by van Breugal et al (1984) in
their regions marked ``Nucleus" and ``F3" sum to 
18.6$\times10^{-15}$ erg cm$^{-2}$ s$^{-1}$, which corresponds to
4378 DN, after an additional 60\% has been added for other lines in F702W.
Thus, it is likely that line emission that is correlated with the
location of the dust is significantly affecting the observed slope and
intercept in Fig. \ref{ext}.  However, the van Breugal flux comes
from an area larger than our dust region, so it is unclear whether
there is really enough emission to produce the entire difference
in slope.
HST observations in line-free filters are
required to clarify the true extinction law in the lane.

We estimated the dust mass by finding the average extinction 
in the dust lane region, excluding knots of emission, 
and 
converting it into V-band extinction using
\beq
A_{V} = A_{\lambda} (a(\lambda) + \frac{b(\lambda)}{R_{V}}) 
\eeq
where a$(\lambda)$ and b$(\lambda)$ are given in
Cardelli \etal (1989), and the effective wavelengths for our
filters are given by Holtzman \etal (1995).  The gas mass was calculated using
\beq
N_{H} = 5.8\times10^{21} \frac{A_V}{R_V}  \rm ~~~~atom ~cm^{-2}
\eeq
which was found empirically by Bohlin \etal (1978), and then summing
the number of atoms using the projected area of the region
(\app 10 h$_{75}^{-2}$ kpc$^2$).  We assume a gas/dust ratio of 100
and  $R_{V}=3.1$.  This procedure gives  a different dust mass
for each filter:
M$_{555}$ = $2.9\times 10^{5}$ h$_{75}^{-2}$ M\solar\ and
M$_{702}$ = $1.9\times 10^{5}$  h$_{75}^{-2}$  M\solar.  
These masses represent lower limits because of the presence of
emission and foreground starlight, and possibly because of a 
non-Galactic extinction law.
M$_{555}$ is more accurate because the F555W band contains less
line emission than F702W.
It is likely that the dust is located near the center of the
galaxy because the nuclear radio source appears to be 
interacting with the dust lane, and because our maximum
extinction is about 0.75 mag.  
Compensating for the foreground light, we find 
M$_{555}$= 6.5$\times$10$^5$ M$_{\odot}$.  This is still 
a lower limit to the dust mass in A1795 because 
of line contamination in F555W,
and because dust may be
concealed in emission filaments (e.g, the {\bf SLF}, Fig. 1)
or in diffuse form.

\section{Discussion}

\begin{center} {\it What is the origin of the dust?} \end{center}

The images clearly show dust in the A1795 CFD despite the harsh environment.
The central temperature from X-ray spectra is 8.5$\times 10^6$ K
(White et al 1994), and a likely core gas density is \app 0.03 cm$^{-3}$ 
(Ge \& Owen 1993).
Using these values, thermal sputtering is expected to remove refractory dust
grains of radius 0.1 $\mu$m in only 6.7$\times 10^{6}$ yr (Draine \& Salpeter
1979a).  Since the lifetime is proportional to the grain radius,
even if the dust were deposited 10$^8$ yr ago, all grains with
radius $a \leq 1.5 \mu$m would already be destroyed.
Clearly, the majority of the dust lane has been
shielded from this hot gas environment.
The extended low-ionization gas and recent star formation near
the dust lane are evidence for a cocoon of gas with T$<<10^7$ K.
This cocoon region may be the working surface of an ``evaporation flow"
(de Jong etal. 1990) where the hot ICM is ionizing gas and sputtering
dust.

A complete sample
of ellipticals from the RSA catalog has been studied by
Goudfrooij et al (1994a; GDHN). They detect dust in 41\% and
ionized gas in 57\%.  They argue for an external
origin for the line+dust systems because the ionized
gas is usually dynamically decoupled from the stellar velocity
field and because internal mass loss cannot account for the quantity
of dust.  Their distribution of masses extends
from 10$^{3.2}$M\solar\ to 10$^{6.3}$M\solar\ with an average near 10$^{4.5}$M\solar,
so the dust mass in A1795 is relatively large.
They measure apparent R$_V$ values which are, on average, lower
than Galactic. They find a tendency for regular
dust morphologies (e.g., smooth rings, disks) to have the 
lowest R$_V$ values, and they conclude that both indicate highly
evolved dust lanes.
It is difficult to draw conclusions from the dust in A1795 because the 
R$_V$ is still undetermined, and its irregular dust morphology
may be caused by radio jet ram pressure rather than by accretion.

NGC 1275 (Perseus) and NGC 4696 (Centaurus) may provide better comparisons
to the CFD in A1795.
They are both centered in cooling flows and contain ionized gas, 
dust with irregular morphology, and powerful
radio sources.   The dust masses are also above average: 
NGC 1275 has 10$^{6.3}$ M\solar\ (Goudfrooij \etal 1994a) and NGC 4696 
has 10$^{5.7}$ M\solar\ (GDHN, SMG).  
The extinction law appears to be indistinguishable from
Galactic in these cases (SMG; GDHN; N{\o}rgaard \etal 1993).
Most of the dust in NGC 1275 is suspected to be a part of 
a system of ionized gas and recent star formation which
has a systematically high velocity with respect to the
nucleus (N{\o}rgaard \etal 1993).
The dust morphology of NGC 4696 is `half-spiral', like A1795,
but does not appear to be interacting with its compact
radio source (SMG).  Thus, the best
explanation for the dust in NGC 1275 and NGC 4696 is a 
recent galaxy merger.
The fact that so many similarities exist between these
three CFDs suggests a connection between the cooling
flow phenomenon and mergers.
One problem with this hypothesis is the expected rarity of
mergers in rich cluster environments (e.g., Merritt
1985).  Also, the CFD in A2029 
contains no dust or ionized gas near its nucleus (McNamara
\& O'Connell 1992).

Although accretion from another galaxy is a likely origin of
the dust lane, it is unlikely that all of the complexities of A1795 can
be explained by one merger.  In particular, the emission filaments
are mostly blue-shifted with respect to the center, they appear
independent of the dust lane, and the line profiles near the central nebula
show that two emission line systems are present (Anton 1993).
This is similar to NGC 1275, whose emission-line filaments appear 
to be independent of the merging, high-velocity  system (N{\o}rgaard \etal 1993).
The widespread blue light (\cite{MO92}) is also difficult to 
explain with a single, recent merger.

\begin{center} {\it What is the source of the blue lobes?}
\end{center}

Our data support the hypothesis that young stars are the
source of the blue lobes in A1795.  Apparent star clusters are now 
resolved at the positions of the lobes by these WFPC2 data.  
A comparison of our smoothed color map to our unsmoothed 
maps show that the bluest
(V-R) ``lobe" occurs where star-cluster-like knots and 
diffuse emission are found.
The northern blue lobe of MO93 
is not apparent in smoothed (V-R), but coincides with regions 
of dust embedded with bluish star clusters.
Very young stars are required for the U-I values, and
so molecular gas may be present, probably with the dust lane
and (dusty) linear filament which contain the lobes.

\begin{center}{\it What role does the radio source play?}
\end{center}

We confirm that the blue lobes occur
along the outer edge of the radio lobes of A1795.
The knotty and diffuse bluish emission which appears to
be the source of the U-band lobes is all located within 1 \h75 kpc
of the radio lobes, in projection (Fig. 3).
In general, bluish star clusters appear to be 
located preferentially near the lobes and jets. 
Thus, our data supports the enhancement of star formation
by the radio source. 
However, the trailed clusters ({\bf T}, Fig. 2) are bluish but 
appear in a region shielded from the
radio source, so not all star formation is radio-induced.
Lobe-induced star formation is feasible because of the
heavy shocking expected at the lobe/IGM interface (Begelman
\& Cioffi 1989).
The radio source may also provide the shock ionization
of gas needed to explain spatial fluctuations in certain line 
ratios (e.g., [OIII]/H$\beta$, Anton 1993).

We also see that dust is located on the outskirts of the radio
lobes.  It is ambiguous whether the radio jets have  shaped
the dust or vice-versa, but an interaction is clear.
This system should provide an interesting opportunity to study
the interaction of radio jets with dusty media.

\acknowledgments

We thank B. McNamara for providing the digitized radio image.
We are grateful to R. Walterbos, and J. O. Burns for engaging 
discussions.  
We utilized the NED Database and the ADS Abstract service for our 
research.  
This work was supported in part by NASA under contract NAS7-918 to JPL.

\clearpage

\figcaption{
Color image of the CFD in A1795 created by assigning F702W to
red, F555W to blue and a combination of the two to green.
North is 55\deg CCW from up. The image is 18\arcs (20.2\h75 kpc)
on a side.
\label{col} }

\figcaption{
Image created by subtracting independent elliptical fits
to the galaxy light from the F702W and F555W images
and then averaging.  The emission features (light) are
emphasized with this technique.
The absorption features (e.g., dust) appear dark.
Labeled features are discussed in the text.
North is to the top and east is to the left.
}

\figcaption{
Overlay of 3.6 cm radio contours (scanned from Ge \& Owen 1993) onto
an extinction map greyscale.  The extinction map was created
by dividing the F702W and F555W images by elliptical models
and then averaging to bring out faint features.
North is to the top.  Absorption features appear white.
}

\figcaption{
Average extinction in F555W and F702W filters for 210
3X3 pixel regions in the dust lane.  The ordinate is
extinction in the F555W band, and the abscissa is
extinction in the F702W band.
The solid line is a fit to points with A$_{702} < 0.3$.
The dashed lines are the limits of extinction expected for a
Galactic extinction law (R$_{V}$=3.1).
Lower limit: a dust sheet of infinite optical depth for varying
depths into the galaxyfor.
Upper limit: a dust screen
positioned in front of all galaxy light (x, the fraction of light in
front of the dust, = 0) with varying
optical depth.  The open circles trace
out the extinction expected for $R_{V}$=3.1 and x=0.5.
The filled circles are for $R_{V}$=1.7 and x=0.5.
}

\clearpage

\setcounter{figure}{3}

%
%
%
  
\begin{figure}[bhp]
\centering \leavevmode
       \epsfxsize=1.0\textwidth
       \epsfbox{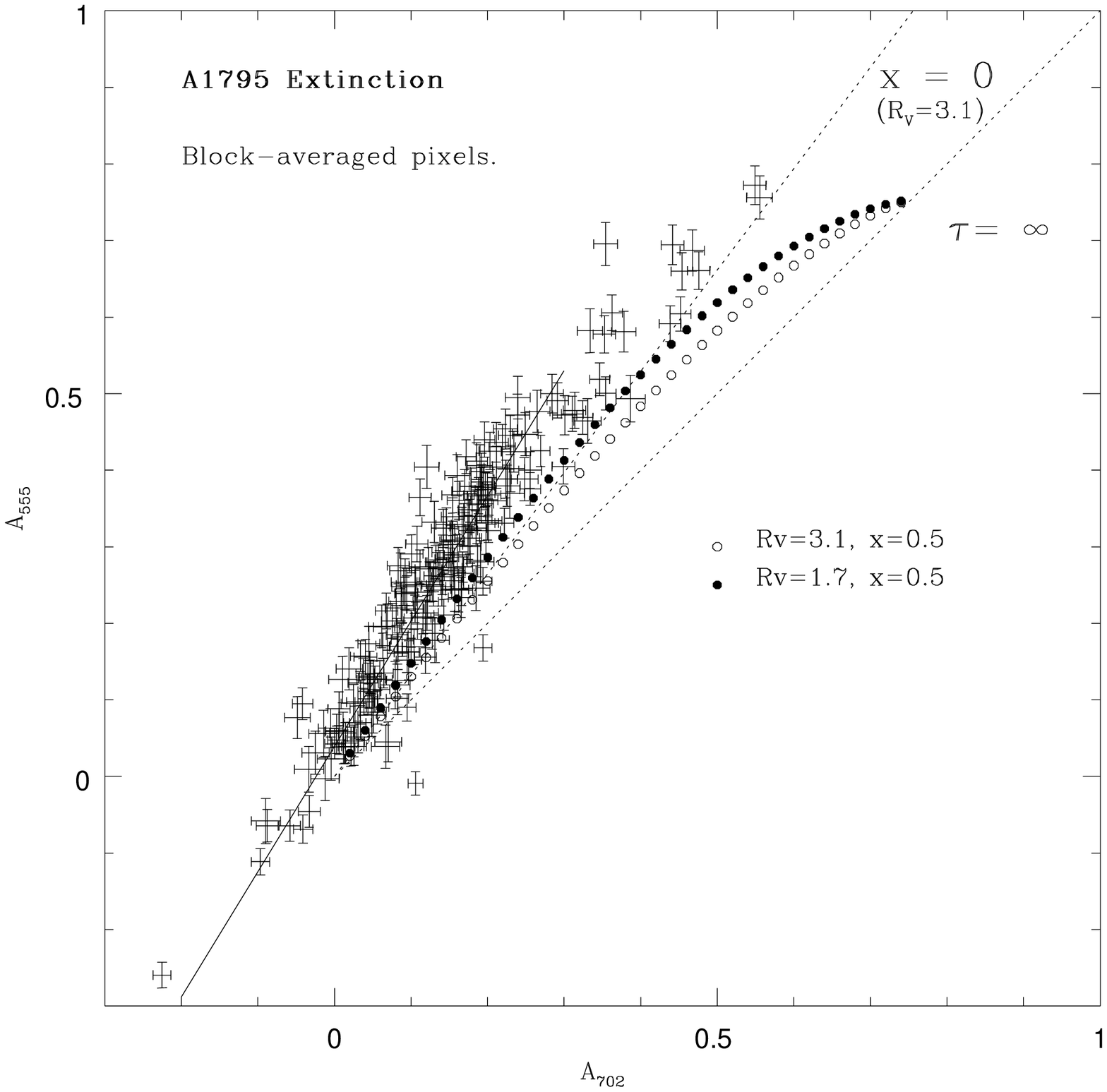}
\caption[Fig. 4]{Figure 4}
\label{ext}  
\end{figure}

\end{document}